\documentclass[aps,prd,reprint,superscriptaddress,floatfix]{revtex4-2}

\usepackage{graphicx}
\usepackage{hyperref}
\usepackage{threeparttable}

\begin{document}

\title{Comment on Transverse Charge Density and the Radius of the Proton}

\author{Benjamin Boone}
\affiliation{Thomas Jefferson National Accelerator Facility, Newport News, VA 23601 USA}
\affiliation{Floyd E. Kellam High School, Virginia Beach, VA 23456 USA }

\author{Michael Chen}
\affiliation{Thomas Jefferson National Accelerator Facility, Newport News, VA 23601 USA}
\affiliation{Hillcrest High School, Midvale, UT 84047 USA}

\author{Kevin Sturm}
\affiliation{Thomas Jefferson National Accelerator Facility, Newport News, VA 23601 USA}
\affiliation{Grafton High School, Yorktown, VA 23692 USA}
\affiliation{Governor's School for Science and Technology, Hampton, VA 23666}

\author{Justin Yoo}
\affiliation{Thomas Jefferson National Accelerator Facility, Newport News, VA 23601 USA}
\affiliation{Grafton High School, Yorktown, VA 23692 USA}
\affiliation{Governor's School for Science and Technology, Hampton, VA 23666}

\author{Douglas Higinbotham}
\affiliation{Thomas Jefferson National Accelerator Facility, Newport News, VA 23601 USA}

\date{\today}

\begin{abstract}
     The charge radius of the proton is typically determined from electron-proton scattering by extracting the proton's electric form factor and then making use of the derivative of that form factor at zero four-momentum transfer.    Unfortunately, experimentally, one cannot measure to zero four-momentum transfer and thus extrapolation is often required.   In the work of Alexander Gramolin and Rebecca Russell, they present a  novel method that does not use the slope and found a radius of the proton that contradicts many other recent results.  Our analysis of their paper discusses some issues with this method and we show that by simply changing the binning of the data and/or including an additional set of data the results change dramatically. 
\end{abstract}

\maketitle

\section{Introduction}

Since early research during the 50s and 60s~\cite{Hofstadter:1955ae, Hand:1963zz}, a variety of different methods have been used to determine the proton charge radius, $r_E$, from electron scattering data producing a wide array of radius values from Hand's low value of 0.805 fm to Borisyuk's high value of 0.912 fm~\cite{Frerejacque:1965ic,Akimov:1972mya,Murphy:1974zz,Borkowski:1974mb,Borkowski:1975ume,Hohler:1976ax,Simon:1980hu,McCord:1991sd,Wong:1994sy,Mergell:1995bf,Rosenfelder:1999cd,SELEX:2001fbx,Sick:2003gm,Kelly:2004hm,Blunden:2005jv,Belushkin:2006qa,PhysRevD.82.113005,Borisyuk:2009mg,PhysRevLett.105.242001,JeffersonLabHallA:2011yyi,Zhan:2011ji,Sick:2012zz,Adamuscin:2012zz,Lorenz:2012tm,Graczyk:2014lba,Lorenz:2014yda,Arrington:2015ria,Lee:2015jqa,PhysRevC.93.065207,PhysRevC.93.055207,PhysRevC.95.035203,PhysRevC.99.044303,PhysRevC.99.055202,Xiong:2019umf,PhysRevC.102.035203,Hayward:2018qij,Mihovilovic:2019jiz,Atac:2020hdq,Lin:2021umk,PhysRevLett.127.092001,PhysRevLett.128.052002}. 

Perhaps the most significant drawback of the electron-proton scattering method of extracting the proton radius 
is that it naively requires experimentalists to make an extrapolation to $Q^2=0$ since
 \begin{equation}
     r_E \equiv \sqrt{-6\frac{dG_E(Q^2)}{dQ^2}\Bigr|_{Q^2=0}}\hspace{2mm},
 \end{equation}
where $G_E$ is the charge form factor of the proton, yet the 
 measurements must be taken at a four momentum transfer ($Q^2 > 0$).
The work of Gramolin and Russell~\cite{Gramolin:2021gln} aims to avoid the issues associated with extrapolation by introducing a novel extraction method relating the proton's charge radius to its transverse charge density. This new method also utilizes new parametrizations of the Dirac and Pauli form factors to model the electron-proton scattering data. 
Gramolin and Russell apply their novel method to the data collected by the A1 Collaboration at the Mainz Microtron MAMI finding a proton radius of $ 0.889 (5)_{\mathrm{stat}}(5)_{\mathrm{syst}}(4)_{\mathrm{model}} \text{ fm,}$ which agrees with the original A1 analysis~\cite{PhysRevLett.105.242001,PhysRevC.90.015206}. 
 In this comment, we replicate their novel method to model and extract $r_E$ from 3 additional data sets, each derived through slight variations of the original A1 set used by the authors. Further, we compare each of these models, in addition to the author's original model, to Sachs form factor ratio data from spin dependent electron scattering asymmetry experiments; this serves as validation of their model using data independent of the original fit, which is absent in the authors' work. 

\section{Different Input Data Sets}

Using Gramolin and Russell's method, we extracted the charge radius of the proton from 4 electron-proton scattering data sets. These four sets were the original A1 Collaboration data \cite{PhysRevC.90.015206}, a rebinned set of the A1 data with corrected and combined data points \cite{Lee:2015jqa}, a data set of the original A1 data combined with a smaller set of high precision, low $Q^2$ cross section data known as the PRad data~\cite{Xiong:2019umf}, and a set of the rebinned A1 data combined with the same PRad data \cite{Lee:2015jqa, Xiong:2019umf}. These data sets will be refered to as the "A1", "Rebinned A1", "A1 + PRad", and "Rebinned + PRad" sets, respectively. All of these data sets are available in the supplementary material to this article. 

In these extractions, we exactly replicated Gramolin and Russell's novel method using a modified version of the PYTHON code they provided and the procedures detailed in section IV of the article. Importantly, note that each of the 3 additional sets constitutes a relatively small deviation from the original A1 set used by the authors; one is merely the same set rebinned, and the number of data points used from the PRad experiment is less than 5\% of the number of points from the A1 Collaboration. Additionally, the original code allows up to the 8th $N$ due to the calculation of the covariance matrix. The Supplementary Material contains the additional data sets that can be used along with a slightly modified version of Gramolin's PYTHON code.

Tables I, II, III, and IV contain the group-wise cross-validation results for different expansion orders $N$; the left side is without regularization, and the right side is--for those $N \geq n$ where $n$ is the order at which $\chi^2_{test}$ was minimized--with optimized regularization, in analogy to Table I from the article. Similarly, Tables IV, V, VI, and VII contain the results from the fits, with optimized regularization, on the full data set (for $N \geq n$), corresponding to Table II from the article.

Running the code on different IDE's and computers results in slight differences for some of the numbers in our tables. These numbers are marked with a dagger ($\dagger$) in the tables. The differences do not create any significant change in other resultant numbers at our precision and do not distort the scientific meaning of our conclusions.
\footnote{The dagger ($\dagger$) signifies slight differences in rounding based on different IDE's and computers.} 

% Gramolin and Russell's original Table I
\begin{table}[ht!]
\begin{threeparttable}
\label{table:gramolin_recreation}
\caption{Our recreation of Gramolin and Russell's first table with the A1 data, with rounding conducted after optimal lambda was chosen, not during.}
\centering
\begingroup
\setlength{\tabcolsep}{6pt} % Default value: 6pt
\renewcommand{\arraystretch}{1.2} % Default value: 1
\begin{tabular}{c c c c c c c c c c c} 

 \hline\hline
  & & \multicolumn{3}{c}{$\lambda=0$} & & \multicolumn{5}{c}{$\lambda > 0$}\\
 \cline{3-5} \cline{7-11} 
 $N$ & & $\chi^2_{\mathrm{train}}$ & & $\chi^2_{\mathrm{test}}$ & & $\lambda$ & & $\chi^2_{\mathrm{train}}$ & & $\chi^2_{\mathrm{test}}$ \\[0.5ex] 
 
 \hline
 
 1 & & 4934 & & 5114 & & & & & & \\
 2 & & 1949 & & 2029 & & & & & & \\
 3 & & 1876 & & 2358 & & & & & & \\
 4 & & 1854 & & 2255 & & & & & & \\
 \textbf{5} & & \textbf{1574} & & \textbf{1682} & & \textbf{0.02} & & \textbf{1574} & & \textbf{1657} \\
 6 & & 1566 & & 1703 & & 0.07 & & 1571 & & 1664 \\
 7 & & 1557 & & 1912 & & 0.18\tnote{$\dagger$} & & 1570 & & 1672 \\
 8 & & 1544 & & 2060 & & 0.36\tnote{$\dagger$} & & 1569 & & 1679 \\ [0.5ex]

 \hline\hline
\end{tabular}
\endgroup
\end{threeparttable}
\end{table}

% Rebinned Table I

\begin{table}[ht!]
\begin{threeparttable}
\label{table:rebinned1}
\caption{Group-wise cross-validation results for the \underline{Rebinned A1} data set before ($\lambda = 0$) and after ($\lambda > 0$) regularization was applied.}
\centering
\begingroup
\setlength{\tabcolsep}{6pt} % Default value: 6pt
\renewcommand{\arraystretch}{1} % Default value: 1
\begin{tabular}{c c c c c c c c c c c} 

 \hline\hline
  & & \multicolumn{3}{c}{$\lambda=0$} & & \multicolumn{5}{c}{$\lambda > 0$}\\
 \cline{3-5} \cline{7-11} 
 $N$ & & $\chi^2_{\mathrm{train}}$ & & $\chi^2_{\mathrm{test}}$ & & $\lambda$ & & $\chi^2_{\mathrm{train}}$ & & $\chi^2_{\mathrm{test}}$ \\[0.5ex] 
 
 \hline
 
 1 & & 1933 & & 2019 & & & & & & \\
 2 & & 642 & & 688 & & & & & & \\
 3 & & 616 & & 903 & & & & & & \\
 4 & & 580 & & 1105 & & & & & & \\
 \textbf{5} & & \textbf{509} & & \textbf{573} & & \textbf{0.01} & & \textbf{510} & & \textbf{553} \\
 6 & & 506 & & 591 & & 0.03 & & 508 & & 557 \\
 7 & & 505 & & 672 & & 0.08 & & 508 & & 563 \\
 8 & & 499 & & 826 & & 0.14\tnote{$\dagger$} & & 507 & & 569 \\ [.5ex]
%   & &  & &  & &  & &  & & \\[0.5ex] 
% 9 & & 499 & & 826 & & 0.14 & & 507 & & 569\\

 \hline\hline
\end{tabular}
\endgroup
\end{threeparttable}
\end{table}

% OG + PRad Table I

\begin{table}[ht!]
\begin{threeparttable}
\label{table:ogprad1}
\caption{Group-wise cross-validation results for the original \underline{A1 + PRad} data set before ($\lambda = 0$) and after ($\lambda > 0$) regularization was applied.}
\centering
\begingroup
\setlength{\tabcolsep}{6pt} % Default value: 6pt
\renewcommand{\arraystretch}{1} % Default value: 1
\begin{tabular}{c c c c c c c c c c c} 

 \hline\hline
  & & \multicolumn{3}{c}{$\lambda=0$} & & \multicolumn{5}{c}{$\lambda > 0$}\\
 \cline{3-5} \cline{7-11} 
 $N$ & & $\chi^2_{\mathrm{train}}$ & & $\chi^2_{\mathrm{test}}$ & & $\lambda$ & & $\chi^2_{\mathrm{train}}$ & & $\chi^2_{\mathrm{test}}$ \\[0.5ex] 
 
 \hline

 1 & & 5618 & & 5206 & & & & & & \\
 2 & & 2284 & & 2125 & & & & & & \\
 3 & & 2227 & & 2493 & & & & & & \\
 4 & & 2201 & & 2399 & & & & & & \\
 5 & & 1969 & & 2699 & & 0.05 & & 1972 & & 1854.7 \\
 \textbf{6} & & \textbf{1901} & & \textbf{1907} & & 0.24 & & 1966 & & 1852.6 \\
 \textbf{7} & & 1897 & & 2165 & & \textbf{0.63} & & \textbf{1961} & & \textbf{1851.9}\tnote{$\dagger$} \\
 8 & & 1859 & & 2527\tnote{$\dagger$} & & 1.23\tnote{$\dagger$} & & 1957 & & 1852.2 \\ 
% 9 & & 1859 & & 2525\tnote{$\dagger$} & & 1.23\tnote{$\dagger$} & & 1957 & & 1852.2 \\ [0.5ex]
% 10 & & 1859 & & 2529 & & 1.23 & & 1957 & & 1852 \\ 
 
 \hline\hline
\end{tabular}
\endgroup
\end{threeparttable}
\end{table}

% Rebinned + PRad Table I
\begin{table}[ht!]
\begin{threeparttable}
\caption{Group-wise cross-validation results for the \underline{Rebinned A1 + PRad} data set before ($\lambda = 0$) and after ($\lambda > 0$) regularization was applied.}
\label{table:rebinnedprad1}
\centering
\begingroup
\setlength{\tabcolsep}{6pt} % Default value: 6pt
\renewcommand{\arraystretch}{1} % Default value: 1
\begin{tabular}{c c c c c c c c c c c} 

 \hline\hline
  & & \multicolumn{3}{c}{$\lambda=0$} & & \multicolumn{5}{c}{$\lambda > 0$}\\
 \cline{3-5} \cline{7-11} 
 $N$ & & $\chi^2_{\mathrm{train}}$ & & $\chi^2_{\mathrm{test}}$ & & $\lambda$ & & $\chi^2_{\mathrm{train}}$ & & $\chi^2_{\mathrm{test}}$ \\[0.5ex] 
 
 \hline

 1 & & 2261 & & 2108 & & & & & & \\
 2 & & 817 & & 777 & & & & & & \\
 3 & & 790 & & 1001 & & & & & & \\
 4 & & 749 & & 1187 & & & & & & \\
 5 & & 698 & & 1313 & & 0.02 & & 698 & & 670.8\\
 \textbf{6} & & \textbf{676} & & \textbf{737} & & \textbf{0.1} & & \textbf{697} & & \textbf{670.7} \\
 7 & & 675 & & 841 & & 0.26 & & 695 & & 672 \\
 8 & & 662 & & 998 & & 0.51 & & 694 & & 674 \\
% 9 & & 662 & & 997\tnote{$\dagger$} & & 0.51 & & 694 & & 674 \\ [0.5ex]
% 10 & & 662 & & 996 & & 0.51 & & 694 & & 674 \\ 
 
 \hline\hline
\end{tabular}
\endgroup
\end{threeparttable}
\end{table}

% ---------- Tables I ^,   Tables II down  ------

% Gramolin and Russell's original Table II
\begin{table}[ht!]
\caption{Gramolin and Russell's Original Table II}
\label{table:og2}
\centering
\begingroup
\setlength{\tabcolsep}{7pt} % Default value: 6pt
\renewcommand{\arraystretch}{1.2} % Default value: 1
\begin{tabular}{c c c c c c} 

 \hline\hline
 $N$ & $\lambda$ & $L$ & $\chi^2$ & $\langle b^2_1 \rangle$ & $r_E$ \\
 & & & & $(\mathrm{GeV}^{-2})$ & (fm) \\ [0.5ex] 
 
 \hline
 
 \textbf{5} & \textbf{0.02} & \textbf{1584} & \textbf{1576} & \textbf{11.49} & \textbf{0.889} \\
 6 & 0.07 & 1580 & 1573 & 11.42 & 0.887\\
 7 & 0.18 & 1578 & 1572 & 11.37 & 0.885\\
 8 & 0.36 & 1577 & 1571 & 11.33 & 0.883 \\[0.5ex]

 \hline\hline
\end{tabular}
\endgroup
\end{table}

% Rebinned table ii

\begin{table}[ht!]
\caption{Fit and extracted radii for the models trained on the full Rebinned A1 data set.}
\label{table:rebinned2}
\centering
\begingroup
\setlength{\tabcolsep}{7pt} % Default value: 6pt
\renewcommand{\arraystretch}{1} % Default value: 1
\begin{tabular}{c c c c c c} 

 \hline\hline
 $N$ & $\lambda$ & $L$ & $\chi^2$ & $\langle b^2_1 \rangle$ & $r_E$ \\
 & & & & $(\mathrm{GeV}^{-2})$ & (fm) \\ [0.5ex] 
 
 \hline
 
 \textbf{5} & \textbf{0.01} & \textbf{514} & \textbf{510} & \textbf{10.88} & \textbf{0.869} \\
 6 & 0.03 & 512 & 509 & 10.84 & 0.867\\
 7 & 0.08 & 511 & 509 & 10.80 & 0.866\\
 8 & 0.14 & 511 & 509 & 10.77 & 0.865 \\ [0.5ex]
% 9 & 0.14 & 511 & 509 & 10.77 & 0.865 \\ [0.5ex] 

 \hline\hline
\end{tabular}
\endgroup
\end{table}

% OG + PRad table ii

\begin{table}[ht!]
\caption{Fit and extracted radii for the models trained on the full A1 + PRad data set.}
\label{table:ogprad2}
\centering
\begingroup
\setlength{\tabcolsep}{7pt} % Default value: 6pt
\renewcommand{\arraystretch}{1} % Default value: 1
\begin{tabular}{c c c c c c} 

 \hline\hline
 $N$ & $\lambda$ & $L$ & $\chi^2$ & $\langle b^2_1 \rangle$ & $r_E$ \\
 & & & & $(\mathrm{GeV}^{-2})$ & (fm) \\ [0.5ex] 
 
 \hline
 
 5 & 0.05 & 1780 & 1766 & 10.67 & 0.861 \\
 6 & 0.24 & 1775 & 1761 & 10.65 & 0.861 \\
 \textbf{7} & \textbf{0.63} & \textbf{1771} & \textbf{1757} & \textbf{10.64} & \textbf{0.860} \\
 8 & 1.23 & 1767 & 1753 & 10.63 & 0.860 \\
% 9 & 1.23 & 1767 & 1753 & 10.63 & 0.860 \\ [0.5ex]
% 10 & 1.23 & 1767 & 1753 & 10.63 & 0.860 \\[0.5ex]

 \hline\hline
\end{tabular}
\endgroup
\end{table}

% Rebinned + PRad table ii

\begin{table}[ht!]
\caption{Fit and extracted radii for the models trained on the full Rebinned A1 + PRad data set.}
\label{table:rebinnedprad2}
\centering
\begingroup
\setlength{\tabcolsep}{7pt} % Default value: 6pt
\renewcommand{\arraystretch}{1} % Default value: 1
\begin{tabular}{c c c c c c} 

 \hline\hline
 $N$ & $\lambda$ & $L$ & $\chi^2$ & $\langle b^2_1 \rangle$ & $r_E$ \\
 & & & & $(\mathrm{GeV}^{-2})$ & (fm) \\ [0.5ex] 
 
 \hline

 5 & 0.02 & 630 & 626 & 10.09 & 0.841 \\
 \textbf{6} & \textbf{0.1} & \textbf{629} & \textbf{624} & \textbf{10.08} & \textbf{0.841} \\
 7 & 0.26 & 628 & 623 & 10.07 & 0.841 \\
 8 & 0.51 & 626 & 622 & 10.06 & 0.841 \\
 9 & 0.51 & 626 & 622 & 10.06 & 0.841 \\ [0.5ex]

 \hline\hline
\end{tabular}

\endgroup
\end{table}

% --------------------------------- END TABLES -----------------------------------------------

As shown in the tables, our best fits were $N=5,\text{ }\lambda=0.02$; $N=5,\text{ }\lambda=0.01$; $N=7,\text{ }\lambda=0.63$; and $N=6,\text{ }\lambda=0.1$; for the A1, Rebinned A1, A1 + PRad, and Rebinned + PRad data sets, respectively. 

The final extraction of $r_E$ using the authors' extraction method yielded
\begin{equation}
    r_E = 0.889(5)_{\text{stat}}(5)_{\text{syst}}(4)_{\text{model}}~\text{ fm, }
\end{equation}
a replication of the original article's result, and
\begin{equation}
    r_E = 0.869 (8)_{\mathrm{stat}}(5)_{\mathrm{syst}}(4)_{\mathrm{model}} \text{ fm,}
\end{equation}
\begin{equation}
    r_E = 0.860 (3)_{\mathrm{stat}}(8)_{\mathrm{syst}}(1)_{\mathrm{model}} \text{ fm, and}
\end{equation}
\begin{equation}
    r_E = 0.841 (4)_{\mathrm{stat}}(11)_{\mathrm{syst}}(0)_{\mathrm{model}} \text{ fm,}
\end{equation}
for the Rebinned A1, A1 + PRad, and Rebinned + PRad data sets, respectively ~\cite{Gramolin:2021gln}.

Thus, the three data sets we fit using the authors' extraction method yielded radii that were all substantially outside the given uncertainty in $r_E$ found using the original data set.

\section{Validation Data}

Gramolin and Russell perform an additional check by extracting the Sachs form factors, $G_E(Q^2)$ and $G_M(Q^2)$, from their model and comparing them to the values found using the effective Rosenbluth separation in Ref.~\cite{PhysRevC.102.035203}. 
Although we do note that the normalizations produced by their model differed from those of the original A1 analysis by less than 0.3\%, it is nevertheless inconsistent to compare the form factors from their model, which has one set of normalizations, with the form factors calculated from the Rosenbluth separation done with a different set of normalizations. 
Notwithstanding this, although consistency with form factors calculated from a Rosenbluth separation of the A1 data confirms that Gramolin and Russell's model does describe the A1 data set (which it was fit to), it does not
%, nor does any other element of the authors' work, 
validate that their model describes the form factors (and thus $r_E$) in reality. To fill this gap, we compared the authors' model with independent validation data, i.e. data not included in the fit. This data comes from a host of polarized electron scattering asymmetry experiments~\cite{Zhan:2011ji,Punjabi:2005wq,Crawford:2006rz,Paolone:2010qc} that measure the form factor ratio, 
\begin{equation}
    \frac{\mu_p G_E(Q^2)}{G_M(Q^2)},
\end{equation}
\noindent where $\mu_p$ is the proton magnetic dipole moment. This data can be found in the Supplementary Material. 

To test the goodness-of-fit of the models on the asymmetry data, we used the $\chi^2$ function
\begin{equation}
    \chi^2 = \sum_{i}\frac{(s_i^{mod} - s_i^{exp})^2}{(\Delta s_i)^2},
\end{equation}
where $s_i^{mod}$ are the models' Sachs form factor ratios, $s_i^{exp}$ are the asymmetry experiments' Sachs form factor ratios, and $\Delta s_i$ are the statistical and systematic uncertainties on the asymmetry ratios added in quadrature. 

\begin{table}[ht!]
\centering
\begingroup
\caption{$\chi^2$ values for each data set tested against the asymmetry data}
\label{table:7}
\setlength{\tabcolsep}{7pt} % Default value: 6pt
\renewcommand{\arraystretch}{1} % Default value: 1
\begin{tabular}{c c c c c} 

 \hline\hline
    & A1 & Rebinned A1 & A1 and & Rebinned A1\\
    &    &             &   PRad &  and PRad \\
   
 \hline

 $\chi^2$ & 179.1 & 208.7 & 169.1 & 192.5 \\ [0.5ex]

 \hline\hline
\end{tabular}
\endgroup
\end{table}

Table VII includes the $\chi^2$ value for each model, including the original A1 fit from the article. 

    \begin{figure*}
    \centering
        \includegraphics[width=\linewidth]{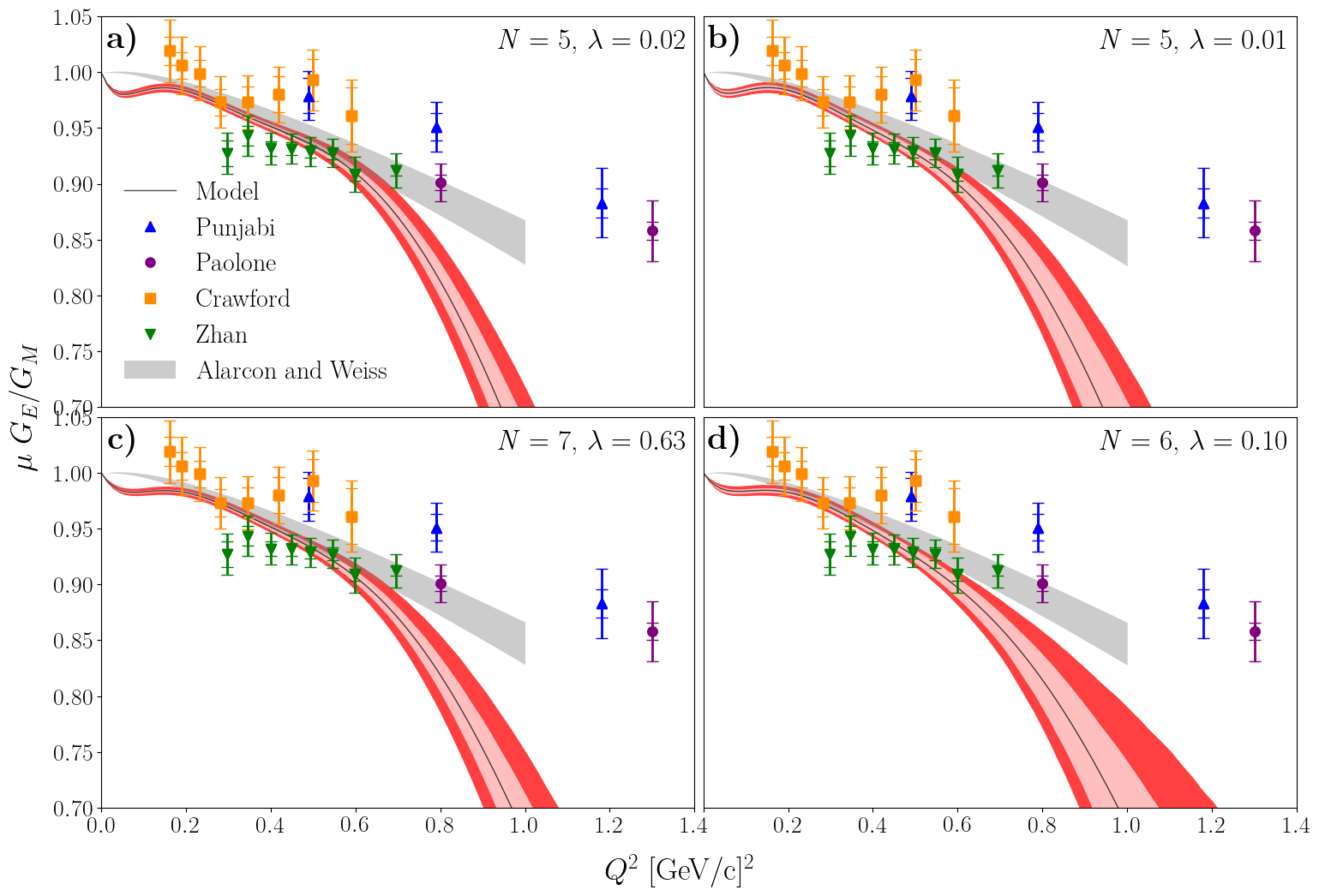}
        \caption{\label{fig:1} Ratio of the Sachs form factors as a function of $Q^2$. Each panel corresponds to a different data set: a) A1, b) Rebinned A1, c) A1 + PRad, d) Rebinned + PRad. Markers: asymmetry data with statistical and systematic uncertainties (consistent across all panels). Grey bands: Alarcon and Weiss's DI$\chi$EFT model fit to the particular data set; the band is the 1$\sigma$ statistical uncertainty, propagated using a Monte Carlo method. The band is consistent across all 4 panels, as DI$\chi$EFT predicted the same model for all sets. Red bands: Gramolin and Russell's best-fit model fit to the particular data set, with the specific order and regularization indicated in the upper right corner of the panel; the black lines are the best-fit lines, the inner, light-red bands are the 68\% statistical confidence intervals and the outer, dark-red bands are the systematic uncertainties added in quadrature. The statistical bands were propagated using standard propagation of error, while the systematic bands were found by following Gramolin and Russell's method outlined in section IV and in their PYTHON code, which itself follows the original A1 analysis; the statistical bands were also calculated using a ratio distribution~\cite{Hinkley:1969} and a Monte Carlo method, both of which were found to be well approximated by the standard propagation. }
    \end{figure*}

In Figure 1,  the form factor ratio is plotted for each data set's best fit model as a function of the four momentum transfer, $Q^2$, together with the asymmetry data that was not included in the fit and is being used here as validation data.
As is clearly exhibited, the authors' model diverges significantly from the asymmetry data at $Q^2$ values larger than ~0.8 $(GeV/c)^2$, indicating that the model does not accurately describe the ratio of the Sachs form factors.

In comparison, consider the DI$\chi$EFT theoretical framework of Alarcon \& Weiss~\cite{Alarcon:2017ivh, Alarcon:2017lhg, Alarcon:2018irp}. DI$\chi$EFT can relate a given proton charge and magnetic radius to a prediction for the $Q^2$ dependence of the form factors, $G_E(Q^2)$ and $G_M(Q^2)$, by utilizing a combination of dispersion analysis and chiral effective field theory~\cite{Alarcon:2020kcz, Alarcon:2018zbz}. Thus, DI$\chi$EFT can serve as a form factor model with 2 parameters, $r_E$ and $r_M$, which can extract the radii values by being fit to electron scattering data. 

In Ref.~\cite{Alarcon:2020kcz}, Alarcon \& Weiss fit their DI$\chi$EFT model to the same set of A1 Collaboration data that Gramolin and Russell used; they found a radius of $r_E = 0.842 \pm 0.002 \text{ (1$\sigma$ fit uncertainty)} ^{+0.005}_{-0.002}$ (full-range theory uncertainty) fm. Unlike Gramolin and Russell's method, when fit to the same 3 additional data sets used in this comment, DI$\chi$EFT found radii consistent with their fit to the original A1 data. Further, as can be seen in Figure 1, Alarcon \& Weiss's model does not suddenly diverge from the asymmetry data for $Q^2 > ~0.8$ in the way that Gramolin and Russell's model does. 

\section{Proton Magnetic Radius}

In addition to the charge radius, one can also extract the magnetic radius from these same fits.  While Gramolin and Russell do not explicitly extract the magnetic radius in their paper, it is straight forward to do and we have done it 
for each data set from their model by calculating $G_M$ from $Q^2$ at very small intervals close to zero and calculating the slope between the points closest to zero, a brute force derivative.  We then use the same equation 1 replacing $G_E$ with $\frac{G_M}{\mu}$ to get the magnetic radius.

We see from those results shown in Table \ref{table:8} that the extracted magnetic radius of the proton is substantially smaller than the respective extracted electric radius. In particular, the value Gramolin and Russell's method obtains from the A1 data differs greatly from the value that the A1 collaboration themselves arrive at, which is 0.777 fm~\cite{PhysRevC.102.035203}, despite them trying to replicate the A1 collaboration's results. 

\begin{table}[ht!]
    \centering
    \caption{The magnetic radius of the proton extracted from each data set using numerically approximated derivatives.}
    \label{table:8}
    \begin{tabular}{l c}
    
    \hline \hline
        Data Set & $r_M$ (fm) \\[0.5ex]
    \hline
    
        Original A1 & 0.7462 \\
        Rebinned A1 & 0.7380 \\
        A1 + PRad & 0.7702 \\
        Rebinned + PRad & 0.7557 \\ [0.5ex]
    \hline \hline
    \end{tabular}
\end{table}

\section{Conclusions}
When using the proton radius extraction procedure from Gramolin and Russel~\cite{Gramolin:2021gln} with different sets of electron scattering data, we find that results are not consistent.  In fact, even a re-binning of the same data results in a vastly different extracted radius.
We also note that while this method exploited the known $Q^2$ limiting, it still failed to extrapolate beyond the data that was included in the fits.   Thus we find that this method is not robust and that the strong conclusions about a particular proton radius that the authors claimed are not justified.

\bibliography{references.bib}

\end{document}